%% file: ICRC2023_template_IceCube.tex
\documentclass[a4paper,11pt]{article}

\usepackage{pos}
\usepackage{multirow}
\RequirePackage{graphicx}
\usepackage{amsmath}
\usepackage{multirow}
\usepackage{lipsum} %package to generate placeholder text in the following

\title{Search for dark matter annihilations in the center of the Earth with IceCube}

\ShortTitle{}

\author{The IceCube Collaboration \\{\normalsize \normalfont(a complete list of authors can be found at the end of the proceedings)}\\}

\emailAdd{Giovanni.Renzi@ulb.be}
\emailAdd{Juan.Antonio.Aguilar.Sánchez@ulb.be}

\abstract{

The nature of Dark Matter remains one of the most important unresolved questions of fundamental physics. Many models, including the Weakly Interacting Massive Particles (WIMPs), assume Dark Matter to be a particle and predict a weak coupling with Standard Model matter. If Dark Matter particles can scatter off nuclei in the vicinity of a massive object, such as a star or a planet, they may lose kinetic energy and become gravitationally trapped in the center of such objects, including Earth. As Dark Matter accumulates in the center of the Earth, self-annihilation of WIMPs into Standard Model particles can result in an excess of neutrinos coming from the center of the Earth and detectable at the IceCube Neutrino Observatory, situated at the geographic South Pole. A search for excess neutrinos from these annihilations has been performed on 10 years of IceCube data, and results have been interpreted in the context of a number of WIMP annihilation channels ($\chi\chi\rightarrow\tau^+\tau^-/W^+W^-/b\overline{b}$) and masses ranging from 10 GeV to 10 TeV. We present the results from this analysis and compare the outcome with previous searches by other experiments. This analysis yields competitive and world-leading results for masses $m_\chi$ > 100 GeV. 
\vspace{4mm}
{\bfseries Corresponding authors:}
Giovanni Renzi$^{1}$, J. A. Aguilar$^{1*}$\\
{$^{1}$ \itshape Universit{\'e} Libre de Bruxelles, Science Faculty CP230, B-1050 Brussels, Belgium}\\[4mm]
$^*$ Presenter

\ConferenceLogo{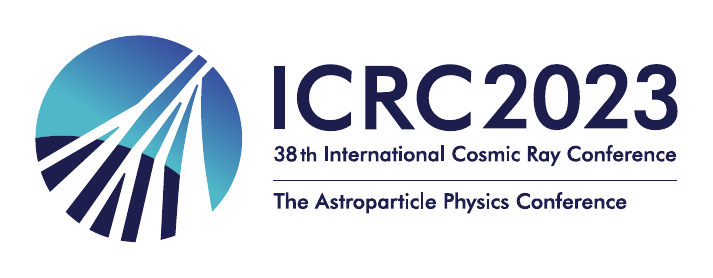}

\FullConference{The 38th International Cosmic Ray Conference (ICRC2023)\\ 26 July -- 3 August, 2023\\ Nagoya, Japan}
}

\begin{document}
\setlength{\bibsep}{1ex}

\maketitle

\section{Introduction}\label{sec1}

Over the last century, an increasing amount of compelling evidence has emerged, indicating that approximately $27\%$ of the Universe and about $85\%$ of its matter content~\cite{Planck:2018vyg} is composed of unknown matter. This elusive form of matter, referred to as {\it dark matter} (DM), must exhibit a low probability of interaction with ordinary matter.

\begin{figure}[h]
    \centering
    \includegraphics[width=.39\textwidth]{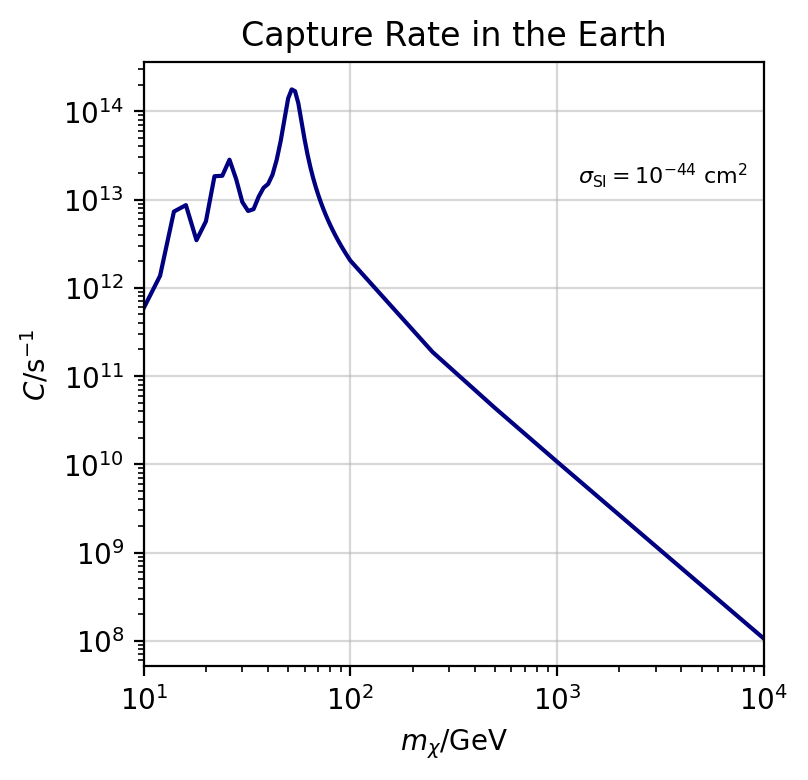}
    \caption{Capture rate as a function of the dark matter particle mass for the spin-independent WIMP-nucleon scattering cross section value \mbox{$\sigma_{\chi N}^{\rm SI}=10^{-44}$ cm$^2$}. The peaks correspond to resonance capture with the most abundant elements on Earth: O, Mg, Si, and Fe. Computed  using DarkSusy \cite{DarkSusy6} \cite{DarkSusyI} \cite{DarkSusy:code}.}
    \label{fig:cap_rate}
\end{figure}

DM particles, denoted $\chi$, can be captured by the gravitational potential of celestial bodies, such as the Sun or the Earth. This capture occurs through repeated scattering interactions with DM particles in orbits intersecting these celestial objects. As a result of these interactions, dark matter particles can lose energy and fall below the gravitational escape velocity. Over time, this will generate an over abundance of dark matter at the center of the celestial body favoring possible process of self-annihilation and therefore reducing the among to DM particles. The process of DM capturing in Earth was first studied in~\cite{1987ApJ...321..571G}. The evolution of number density of DM particles, $N_\chi$, will follow the equation that describes the interplay between capturing and annihilation as:

\begin{equation}
    \label{eq:dm_rate}
    \frac{{\rm d} N_{\chi}}{{\rm d}t} = C_{\rm C}-C_{\rm A}N_{\chi}^2-C_{\rm E}N_{\chi}.
\end{equation}
In this equation $C_{\rm C}$ refers to the capture rate and $C_{\rm A}N_{\chi}^2=\Gamma_{\rm A}$ is the annihilation rate. The third process, represented by $C_{\rm E}N_\chi$, is the so-called {\it evaporation}, which describes the process in which after scattering the DM particles gains kinetic energy above the escape velocity and therefore escapes. Evaporation can be ignored for dark matter masses above $\sim 12\;\rm{GeV}$~\cite{evaporation:2021}. The capturing rate, depends on the local DM density, the scattering cross-section, and the chemical composition of the target body. The local halo DM density cannot be stated with certainty and is expected to be in the range $0.1 < \rho_{\rm DM}/{\rm GeV\:cm}^{-3} < 0.5$ \cite{Necib:2019InferredEF}. The value assumed conventionally and used for the analysis presented in this work is \mbox{$\rho_{\rm DM}=0.3$ GeV cm$^{-3}$}. Figure~\ref{fig:cap_rate} shows the Earth's capturing rate assuming the Earth's composition in~\cite{Earth_composition} as function of the DM mass.

 The peaks in the capturing rate are due to resonance capture with the most abundant elements on the Earth. Most abundant elements on
Earth do not carry spin hence the most relevant scattering process is the spin-independent dark matter-nucleon scattering, $\sigma_{\chi N}^{\rm SI}$.

As can be seen in Eq.~\eqref{eq:dm_rate}, the annihilation rate depends on the dark matter number density squared, as two dark matter particles are needed for annihilation, as well as the self-annihilation cross-section, usually expressed as a velocity averaged cross-section,
$\langle\sigma_{\rm A}v\rangle$. Equation~\eqref{eq:dm_rate} can be solved and has a solution that can be expressed as:
\begin{equation}
    \label{eq:dm_rate_sol}
    \Gamma_{\rm A}(t) = \frac{1}{2}C_{\rm C}\tanh^2\left(\frac{t}{\tau}\right) ,
\end{equation}
where $\tau=(C_{\rm C}C_{\rm A})^{-1/2}$ is the time scale for the capture and annihilation processes to reach equilibrium. The equilibrium condition is fulfilled when $t \geq \tau$, so that Eq.~\eqref{eq:dm_rate_sol} becomes $C_{\rm C}=\frac{1}{2}\Gamma_{\rm A}$. In this case the annihilation rate reaches its maximum. For typical value of the thermal relic annihilation cross section the Earth is far from equilibrium ($t_\oplus \ll \tau$). However, enhancements in the annihilation cross-section such as a nearby clump of DM or a low-velocity Sommerfeld effect, will increase the annihilation rate for the Earth while not for the Sun which is, likely, already in hydrostatic equilibrium~\cite{Delaunay_2009}. Due to its low escape velocity, Earth capturing is very sensitive to the low tail of the DM velocity distribution. However the low speed distribution is not precisely known, and deviations from the standard halo model could significantly enhance the capturing rate~\cite{BRUCH2009250}.

As DM annihilates, it will produce stable particles, such $\gamma$-rays and neutrinos. Given the density of the Earth, only neutrinos can reach the surface and interact in the vicinity of a neutrino telescope. This constitutes the base of the detection principle for indirect searches of dark matter from the center of the Earth.

\section{The IceCube Neutrino Observatory}
\label{sec:icecube}
The IceCube Neutrino Observatory~\cite{IceCube:2016a} is a one-cubic-kilometer neutrino telescope situated at the geographic South Pole. The detector is composed of 5160 photo-multiplier units (DOMs) deployed on 86 strings and is buried in the Antarctic ice-cap between the depths of 1,450 m and 2,450 m. IceCube detects Cherenkov light induced by the passage of superluminal charged leptons and hadrons produced in neutrino interactions in ice. The number of Cherenkov photons and their arrival time is used to reconstruct the direction and energy information of the incoming neutrino. Based on its geometry, IceCube can detect neutrinos with energy ranging from a few GeV to a few PeV. A denser sub-array, \textit{DeepCore} \cite{IceCube:2016a}, is placed between 2100 m and 2450 m of depth and extends the detection threshold to lower energy neutrinos, between 10 GeV and 100 GeV.

\section{Data Selection}

Neutrinos produced as by-products of the DM self-annihilation at the center of the Earth, will reach the IceCube detector with vertical up-going directions. Hence, the data selection is designed and optimized to keep vertical tracks. The first step of the processing focuses on reducing the muon atmospheric background events which arrive to the detector with downward going directions. Once data is reduced to an manageable level, we applied a random forest of boosted decision trees (BDTs) to discriminate mis-reconstructed muon events from neutrino events. A random forest assigns to each event a score ranging from -1 (most background-like) to +1 (most signal-like). At this point, we can observe differences between the highest and lowest energy signal events: low-energy events appear more like cascade events, mostly comprised in DeepCore; high-energy events consist of kilometer-long tracks. In a intermediate step, a higher-performing but computationally intensive energy reconstruction is applied. The last selection based on the random forest classification is optimized to maximize the sensitivity of the analysis. Two BDT are applied, one which is tuned to select high energy signal, and another one optimized for low energy dark matter masses.

\begin{figure*}
    %\centering
    \begin{minipage}{0.49\textwidth}
    \includegraphics[width=\textwidth]{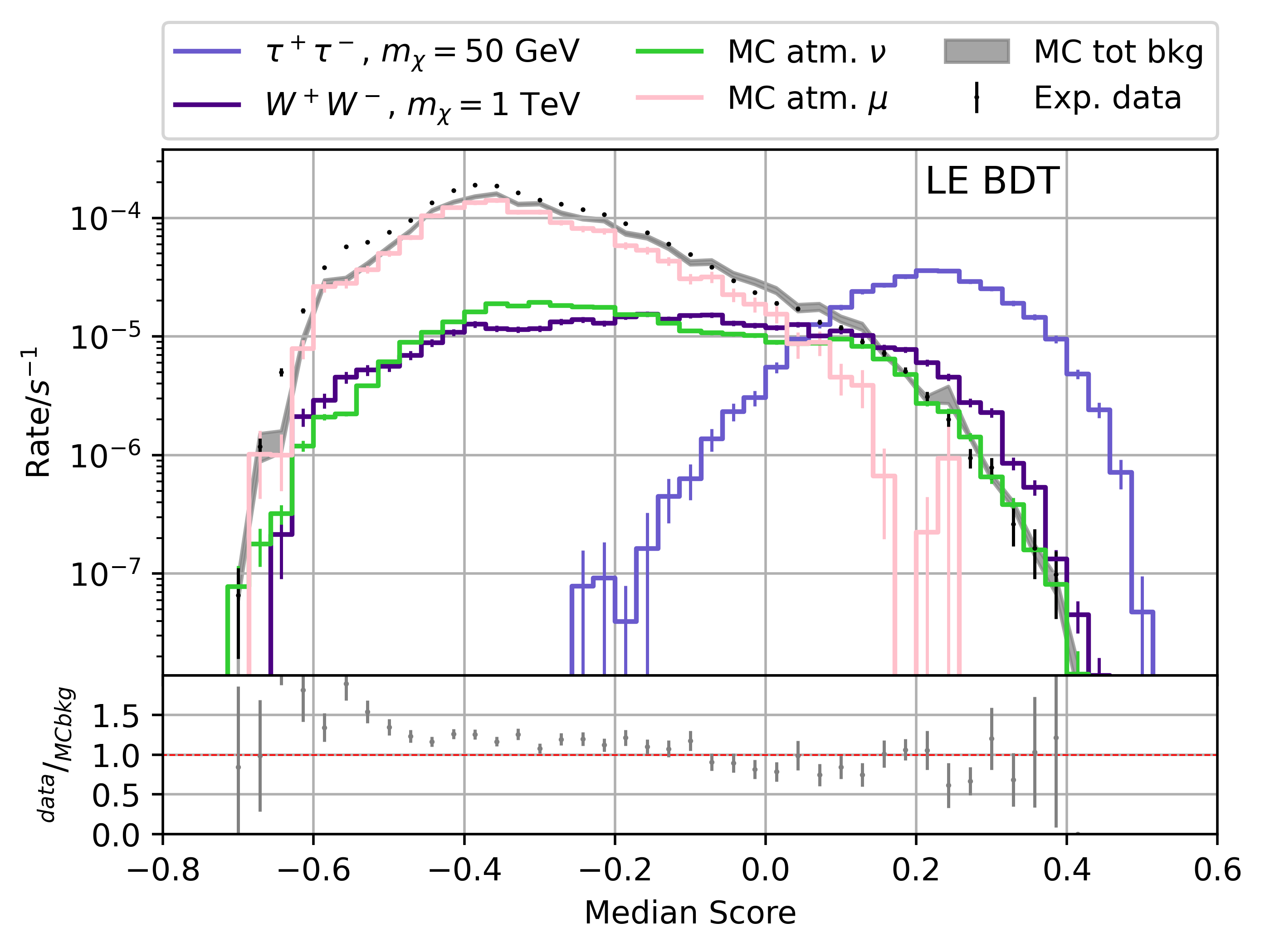}
    \end{minipage}
    \begin{minipage}{0.49\textwidth}
    \includegraphics[width=\textwidth]{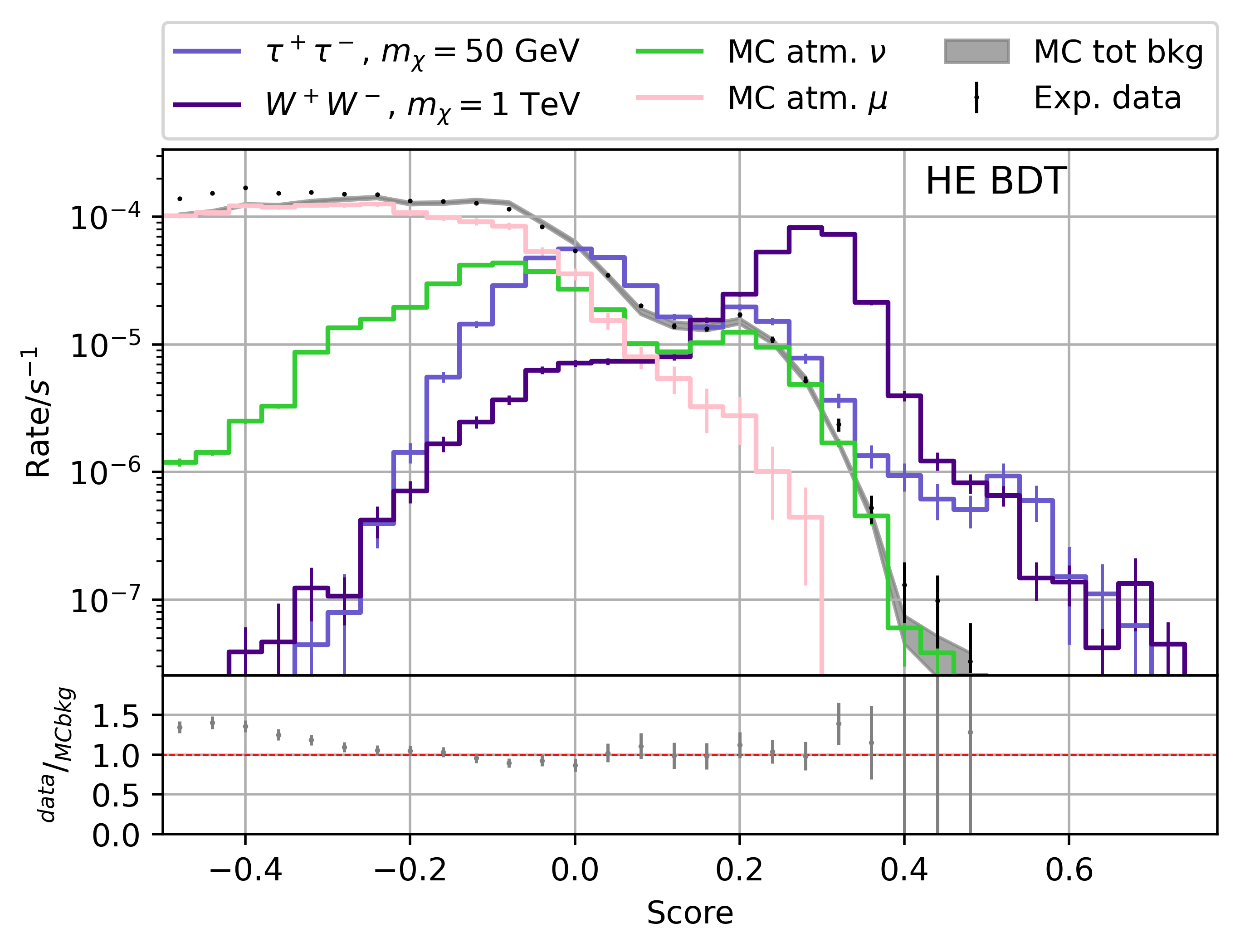}
    \end{minipage}
    \caption{Score distributions for the LE (left) and HE (right) random forests. The baseline LE and HE signal configurations are in light purple and dark purple, respectively. Atmospheric neutrinos and muons are in green and pink respectively. The grey band indicates the total background estimation while the black dots represent the data verification sample.}
    \label{fig:bdt}
\end{figure*}

Figure~\ref{fig:bdt} shows the differences in the score distributions for the LE and HE baseline signal configurations for the two random forests.\\

In this analysis we used 3,619 days of data taking over ten detector seasons from May 2011 to May 2020. A subset of 353 days, taken sparsely over the ten seasons, have been used as a verification dataset along the event selection process. The verification dataset has been omitted from the analysis, leaving a final of 3,266 available days of data.

\section{Background and Signal Simulation}
Neutrinos from the center of the Earth have an unique direction in IceCube's local coordinate system. There is no other direction in the sky in which IceCube has the same detector response, therefore background estimation techniques, like {\it on-source}/{\it off-source} methods, are not possible. This includes the typical method used in searches for point sources in neutrinos telescopes called right-ascension scrambling, which makes use of the uniform response of the detector in right-ascension due to Earth's rotation. For this reason, this analysis relies on MonteCarlo simulation to estimate the background at a perfect upward going direction. The atmospheric muons generated by cosmic-ray showers are simulated using CORSIKA~\cite{corsika:98}. The diffuse atmospheric, prompt~\cite{Sarcevic:PhysRevD.78.043005} and astrophysical neutrinos~\cite{IceCubeDiffuse:2021uhz} are simulated using two different neutrino generators: GENIE \cite{GENIEAndreopoulos:2015wxa} for energies up to 100 GeV and NuGen \cite{ANIS_GAZIZOV:2005203} beyond. The reason for these two simulations is due to the different implementation and energy regime of the neutrino nucleon interaction, with NuGen covering the deep inelastic scattering and GENIE the quasi-elastic interaction. After generating interaction vertexes, all charge particles are then propagated through the detector and their induced Cherenkov light production and detection also simulated.

Neutrinos from the annihilation of DM at the center of the Earth are simulated using the WIMPsim~\cite{WimpSim:code}. Re-weighting of the isotropic simulation from GENIE and NuGen becomes impractical again due to the unique direction in local coordinates of the DM signal. Neutrinos from WIMPSim are generated as coming from the center of the Earth ($\theta = 180^{\circ}$) and their interaction products are propagated and simulated in the same fashion as astrophysical and atmospheric neutrinos. 

We produced simulations over DM particle mass values ranging  from 10 GeV to 10 TeV in three annihilation channels: \mbox{$\chi\chi\rightarrow\tau^+\tau^-$/$W^+W^-$/$b\bar{b}$}. All the simulated mass-channel configurations are resumed in Table~\ref{table:wimpsim}. We chose two baseline signal configurations, a low-energy one for $\chi\chi\rightarrow\tau^+\tau^-$, $m_{\chi}=50$ GeV, and a high-energy one for $\chi\chi\rightarrow W^+W^-$, $m_{\chi}=1$ TeV. As the event selection selects track-like events, only muon neutrinos are considered for these simulations.

\begin{table}
\centering
    \caption{Summary of WIMPs simulation scenarios produced with WimpSim.}
    \label{table:wimpsim}
    \begin{tabular}{c|c}
        Channel & Masses \\
        \hline
        \multirow{3}{*}{$\chi\chi\rightarrow\tau^+\tau^-$} & [10, 20, 35, 50] GeV\\
        %\multirow{3}{*}{$\chi\chi\rightarrow\tau^+\tau^-$} 
         & [100, 250, 500] GeV\\
         & [1, 3, 5, 10] TeV\\
         \hline
         \multirow{2}{*}{$\chi\chi\rightarrow W^+W^-$} & [100, 250, 500] GeV\\
         & [1, 3, 5, 10] TeV\\
         \hline
         \multirow{3}{*}{$\chi\chi\rightarrow b\bar{b}$} & [35, 50] GeV\\
         & [100, 250, 500] GeV\\
         & [1, 3, 5, 10] TeV\\
    \end{tabular}
\end{table}

\section{Analysis Method}

\begin{figure*}
\centering
    \begin{minipage}{0.49\textwidth}
        \includegraphics[width=\textwidth]{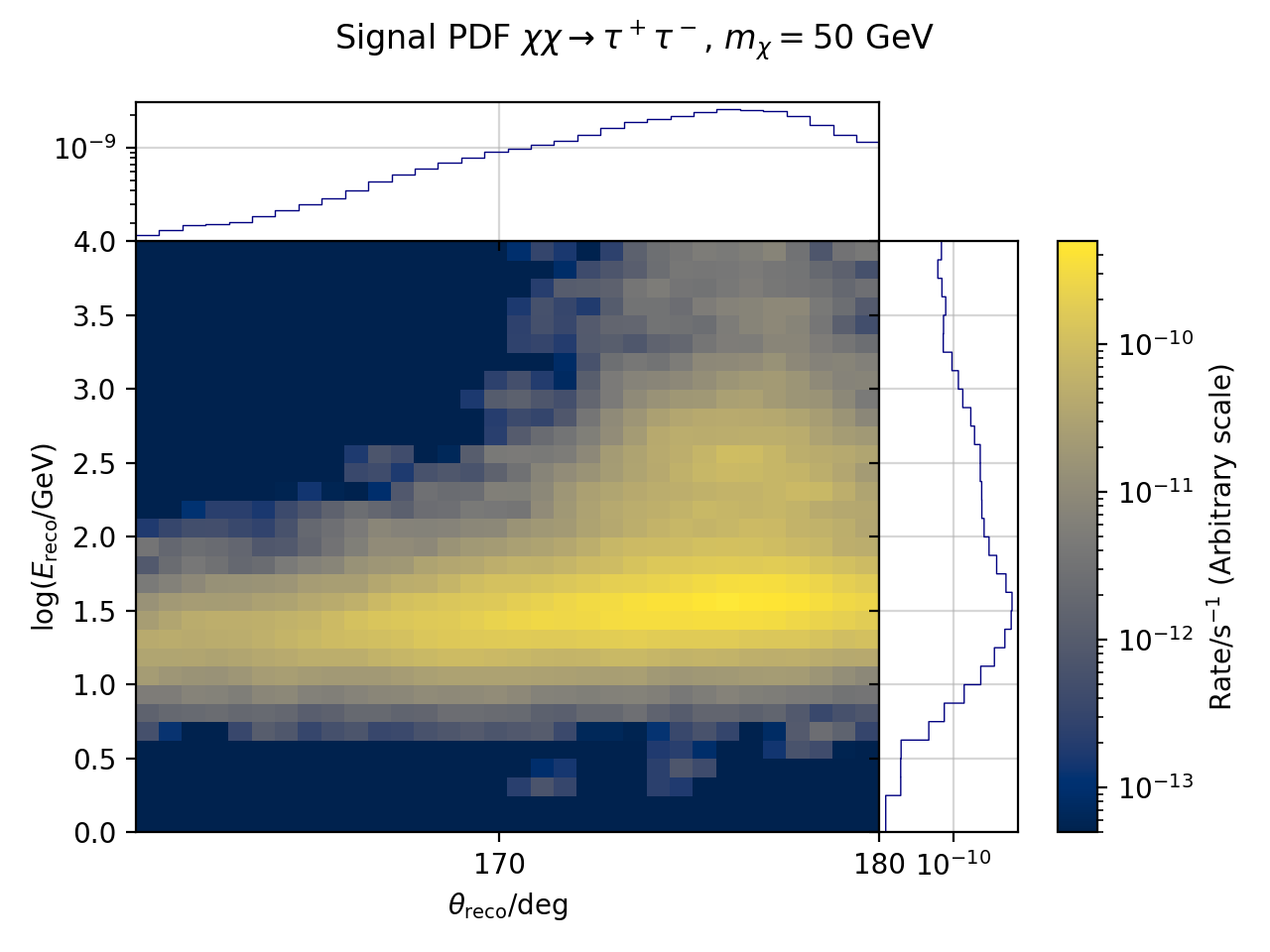}
    \end{minipage}
    \begin{minipage}{0.49\textwidth}
        \includegraphics[width=\textwidth]{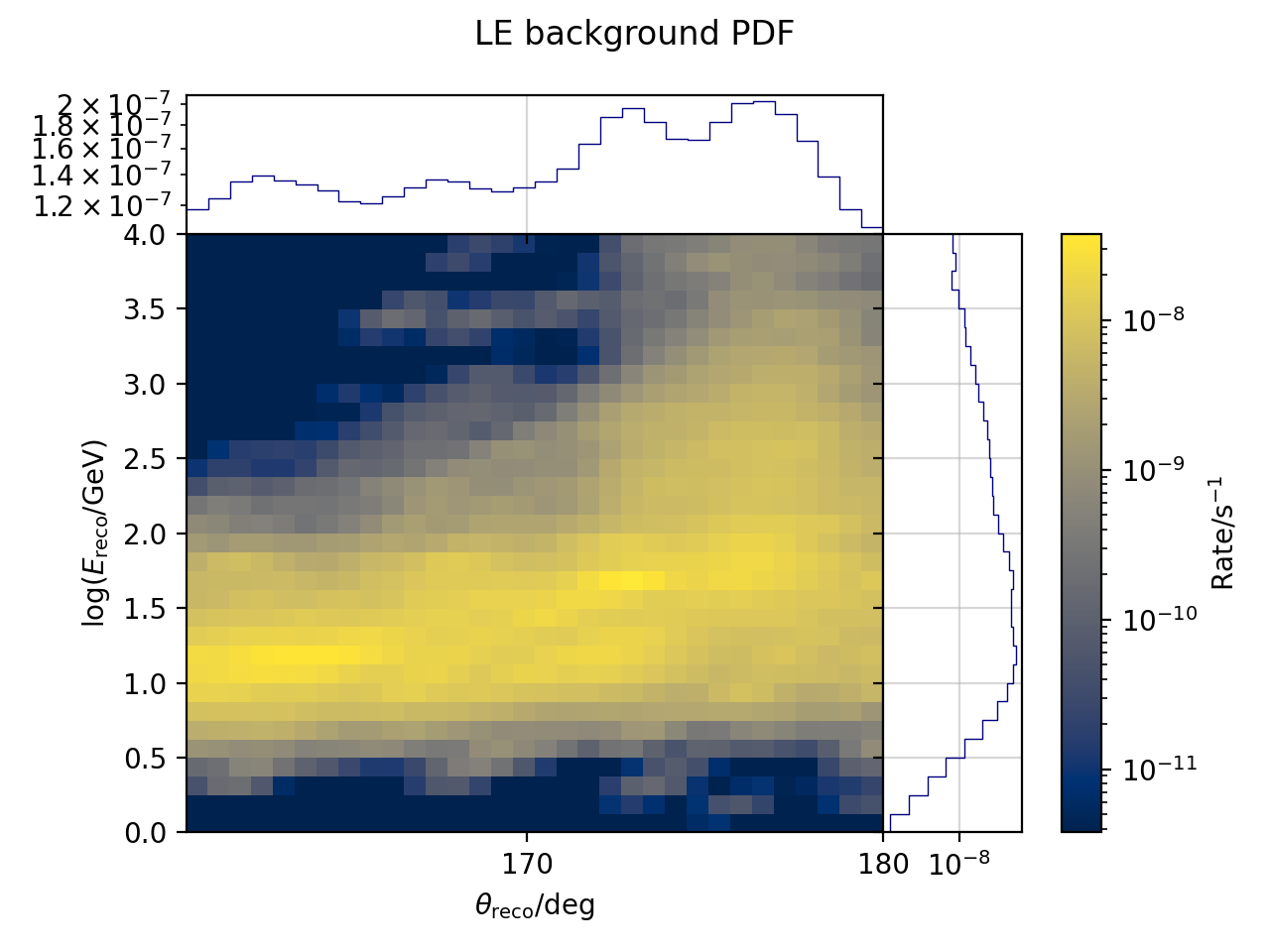}
    \end{minipage}
    \caption{PDFs for the LE analysis. The zenith angle, $\theta_{\rm reco}$, is on the $x$-axis while the log-energy,
    $\log(E_{\rm reco}$, is on the $y$-axis. The LE signal baseline and the atmospheric background are shown on the left and on the right, respectively.}
    \label{fig:le_pdf}
\end{figure*}

\begin{figure*}
\centering
    \begin{minipage}{0.49\textwidth}
        \includegraphics[width=\textwidth]{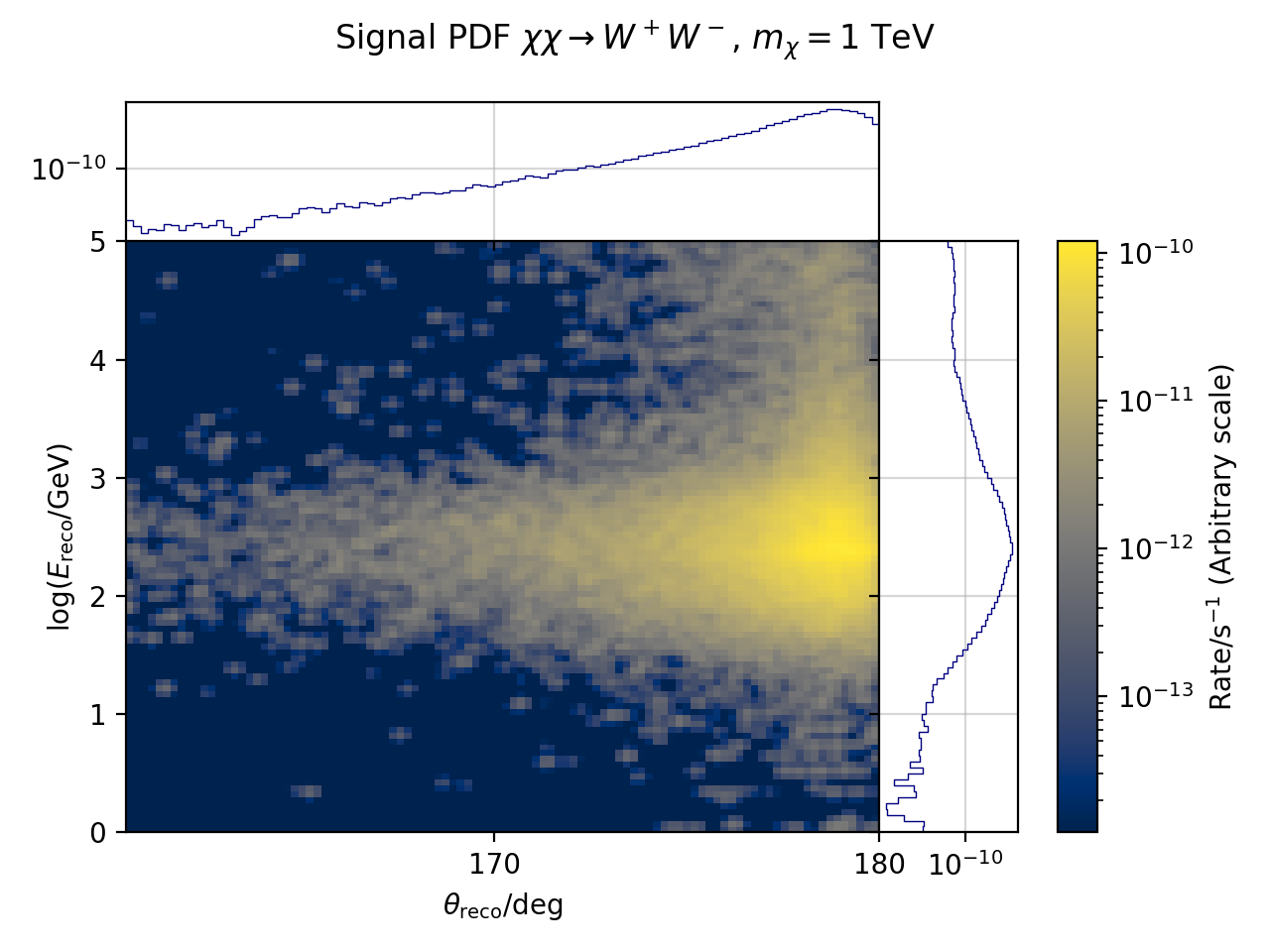}
    \end{minipage}
    \begin{minipage}{0.49\textwidth}
        \includegraphics[width=\textwidth]{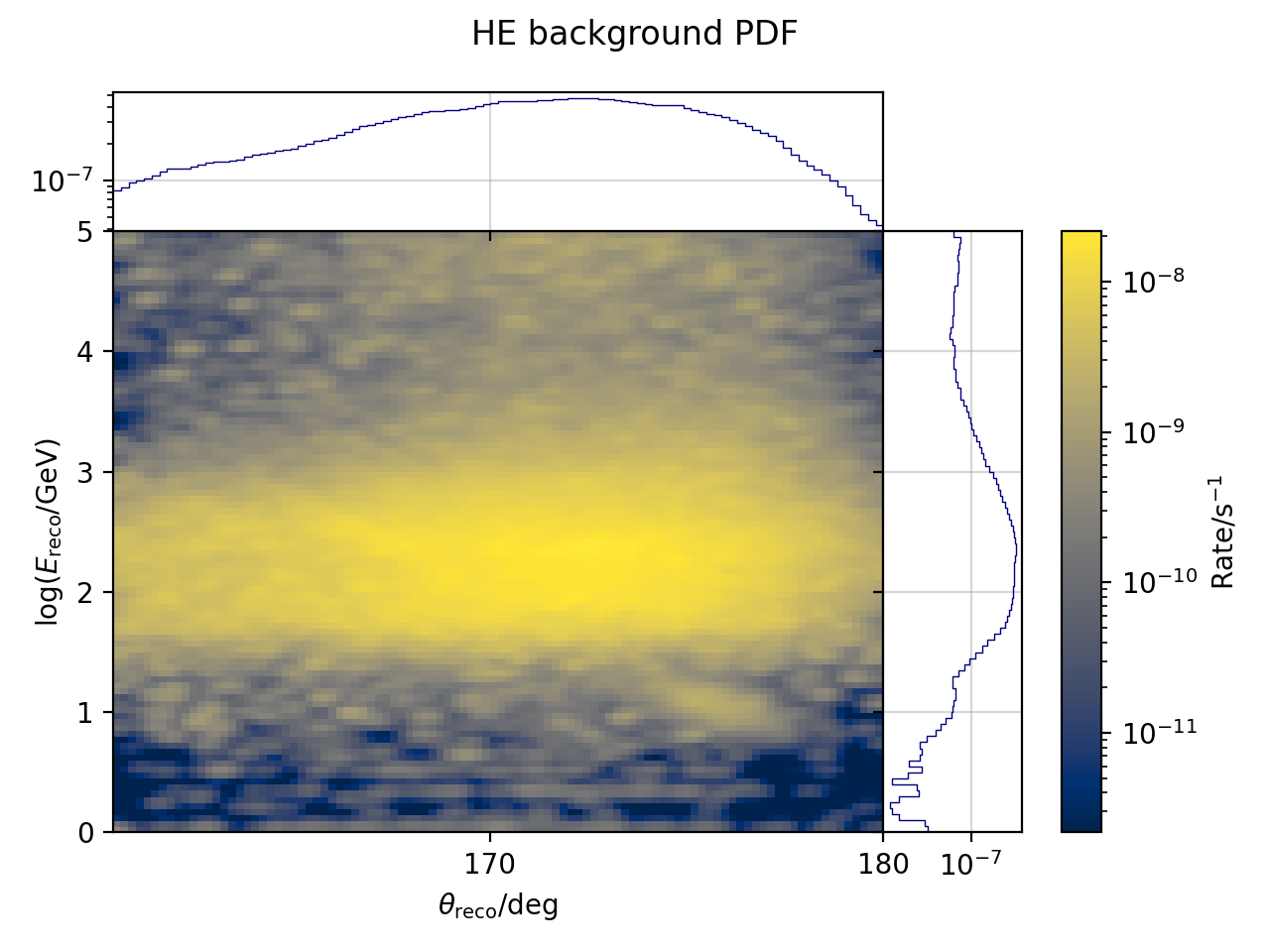}
    \end{minipage}
    \caption{PDFs for the HE analysis. The zenith angle, $\theta_{\rm reco}$, is on the $x$-axis while the log-energy,
    $\log E_{\rm reco}$, is on the $y$-axis. The LE signal baseline and the atmospheric background are shown on the left and on the right, respectively.}
    \label{fig:he_pdf}
\end{figure*}

We used a binned Poisson likelihood method using as observables the reconstructed energy of the event and their reconstructed zenith angle.  The expected signal from DM will arrive from almost vertical directions ($\theta_{rec} \sim 180^{\circ}$). We produce binned probability density functions (PDFs) over these two observables based on the final distributions obtained in the event selection process after applying a kernel density estimation (KDE) technique to cope with statistical fluctuations due to low statistics.  For the LE analysis, we include events with energy in the range $1\; {\rm GeV} <E_{\rm reco}<10^4 {\rm GeV}$. We found the optimised number of bins at $32\times32$, for a total of $1024$ bins. For the HE analysis, the energy range (in GeV) is $1<E_{\rm reco}<10^5$ and the optimised bins grid is $100\times100$ for a total of $10^4$ bins. For both analyses, the zenith angle range is $160^{\circ} < \theta_{\rm reco}<180^{\circ}$.
The PDFs for the signal and atmospheric background are shown in Figs.~\ref{fig:le_pdf} and \ref{fig:he_pdf} for LE and HE, respectively.

The expected fraction of events in the $i$-th bin, $\lambda_i$, can be expressed as:

\begin{equation}
    \label{eq:model_base}
    \lambda_i = \xi S_i + (1-\xi)B_i,
\end{equation}

where $\xi$ is the signal fraction and the parameter against with the likelihood is minimized. The background density $B_i$ is composed, likewise, of many different components: atmospheric neutrinos and muons, astrophysical neutrinos and prompt neutrinos. Their relative normalization are kept as nuisance parameters in the minimization and expressed as:

\begin{equation}
    \label{eq:model}
    \lambda_i = \xi S_i + \sum_{j=1}^{n-1}\prod_{k=0}^{j-1}(1-n_k)n_j B^j_i + \prod_{k=0}^{n-1}(1-n_k)B^n_i,
\end{equation}

where $n_k$ refers to the different background event fractions which can be encompassed by $\overrightarrow{\eta}=(n_1, ...,n_{n-1})$ and $k_i$ being the observed number of events in the $i$-th bin. We then search the parameters, $\xi$, and $\eta$ that minimize the likelihood:
\begin{equation}
    \label{eq:likelihood}
    -\log{\mathcal{L}(\xi,\overrightarrow{\eta})} = \sum_i^{N_{bins}}(-k_i\log\lambda_i+\lambda_i).
\end{equation}

In order to perform an hypothesis testing, we define a test-statistics as the likelihood ratio:
\begin{equation}
\label{eq:ts}
    t_\xi = 2\log\frac{\mathcal{L}(\xi,\hat{\hat{\overrightarrow{\eta}}})}{\mathcal{L}(\hat{\xi},\hat{\overrightarrow{\eta}})} = 2(\log\mathcal{L}(\xi,\hat{\hat{\overrightarrow{\eta}}})-\log\mathcal{L}(\hat{\xi},\hat{\overrightarrow{\eta}})),
\end{equation}
\noindent where $\hat{\xi}$, $\hat{\eta}$ are the estimated parameters that minimize the likelihood, $\mathcal{L}(\hat{\xi},\hat{\overrightarrow{\eta}})$, and $\hat{\hat{\eta}}$ is the value of $\eta$ that minimizes the likelihood for any given value of $\xi$. To claim a discovery we estimate the agreement with respect to the null hypothesis $\xi = 0$, and so the test-statistics is defined as $t_0$. Assuming Wilk's theorem~\cite{Wilks, Wilks2} and that the distribution of $t_0$ under sample of null-hypothesis follows a $\chi^2$ distribution, we can estimate the p-value in terms of number of sigmas $z$-score as $ z-{\rm score} =\sqrt{t_0}$.

\section{Results and Conclusions}

We found no significant excess over all channels and masses tested. The highest pre-trial significance belongs to the HE analysis at $1.94\sigma$ for the channel $\chi\chi\rightarrow b\bar{b}$ and mass \mbox{$m_{\chi}=250$ GeV}. The trial factor correction gives a post-trial significance of $1.06\sigma$. Upper limits in the signal fraction of events $\xi$, can be converted into upper limits on the spin-independent DM-nucleon cross-section $\sigma_{\chi N}^{\rm SI}$ after assuming a value on the DM self-annihilation cross-section.

\begin{figure}
    \centering
    \includegraphics[width=.49\textwidth]{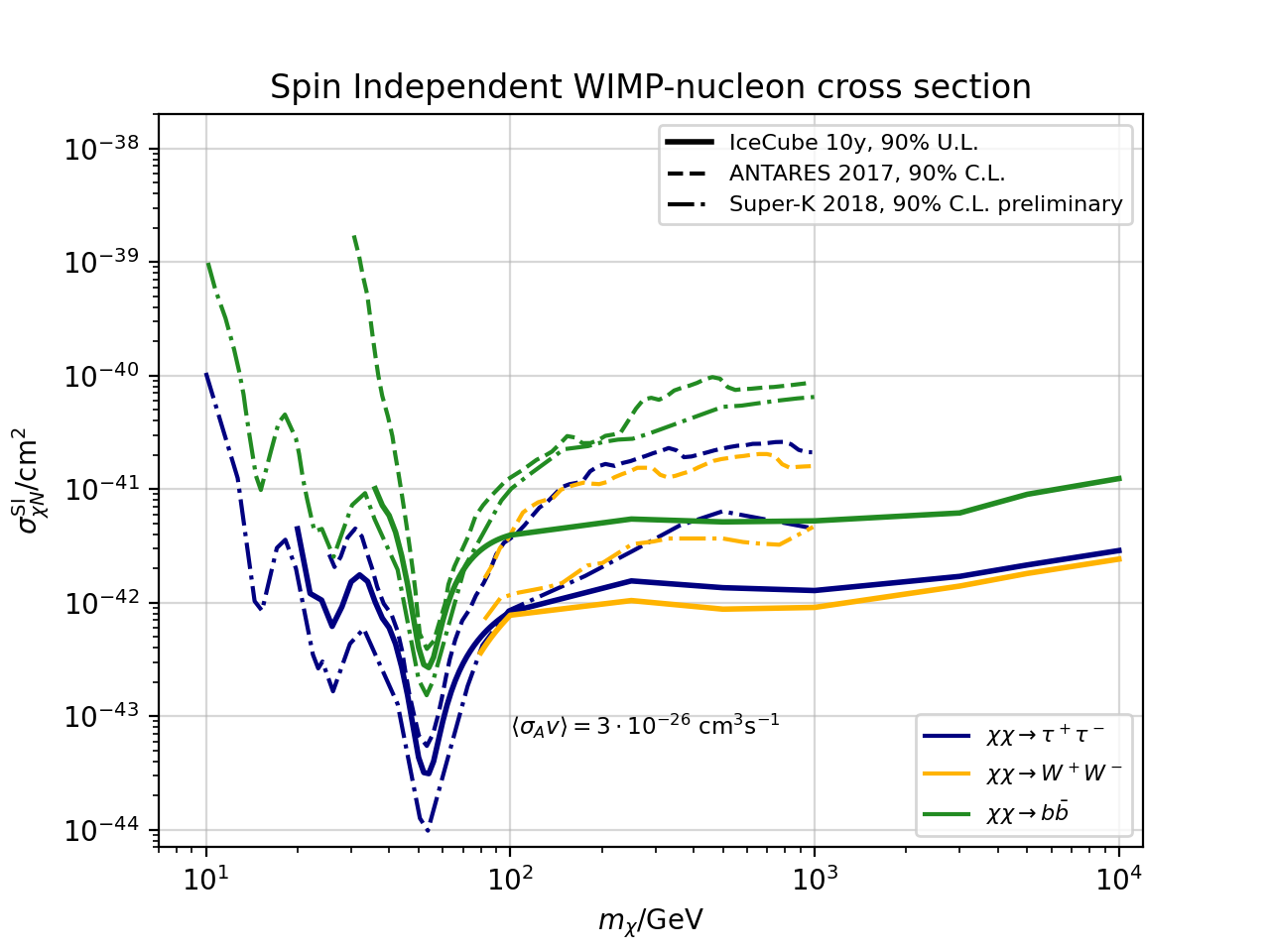}
     \includegraphics[width=.49\textwidth]{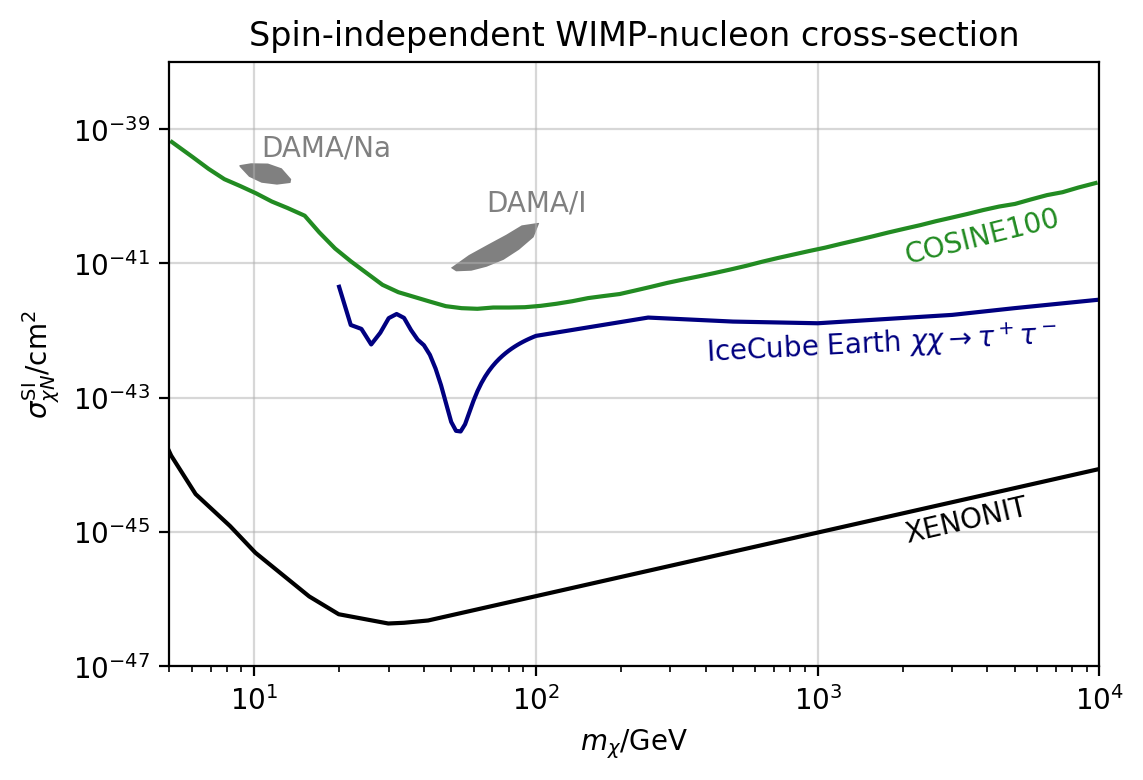}
    \caption{Left: Spin-indenpendent DM-nucleon cross-section $90\%$ C.L. upper limits. Limits in bold are from this analysis, ANTARES (dashed) and Super-Kamiokande (dot-dashed) limits are included. The colour code identifies the annihilation channels, blue for annihilation into $\tau^+\tau^-$, yellow for $W^+W^-$ and green for $b\bar{b}$. Right: Upper limits for this analysis (blue) for the $\chi\chi\rightarrow \tau^+\tau^-$ annihilation channel compared to direct detection upper limits from the crystal experiments DAMA/LIBRA \cite{DAMA:2018} (grey areas) and COSINE100 \cite{COSINE100:2019} (green) and from XENON1T \cite{XENON3:2019} (black).}
    \label{fig:xsec_limits}
\end{figure}

The $90\%$ C.L. upper limits for $\sigma_{\chi N}^{\rm SI}$ are presented in Fig.~\ref{fig:xsec_limits}, compared to results from the other neutrino telescopes ANTARES \cite{ANTARES_Earth:2016} and Super-Kamiokande \cite{SuperK_DM:2020}. 

Although optimized for signals from WIMP-like particles, the results from this analysis can also be re-casted to set limits on different DM models. An example can be found in \cite{Renzi:2022}, where limits on the coupling constant of the effective field theory of dark matter~\cite{Catena_2017} were computed.

With the installation of the IceCube Upgrade~\cite{Ishihara:2019aao} a significant improvement in IceCube's sensitivity can be expected, especially in the low-mass region.

% Bibtex references:
\bibliographystyle{ICRC}
\bibliography{references}

% Alternatively, you can include references by hand:
%\begin{thebibliography}{99}
%\bibitem{...}
%
%\end{thebibliography}

\clearpage

%The following list of authors, affiliations and funding agencies will be updated at the day of submission. The following template is a placeholder generated via https://authorlist.icecube.wisc.edu/icecube on March 25, 2023 and will be updated.
\input{authorlist_IceCube.tex}

\end{document}

%% file: authorlist_IceCube.tex
\section*{Full Author List: IceCube Collaboration}

\scriptsize
\noindent
R. Abbasi$^{17}$,
M. Ackermann$^{63}$,
J. Adams$^{18}$,
S. K. Agarwalla$^{40,\: 64}$,
J. A. Aguilar$^{12}$,
M. Ahlers$^{22}$,
J.M. Alameddine$^{23}$,
N. M. Amin$^{44}$,
K. Andeen$^{42}$,
G. Anton$^{26}$
% ...
\\
\\
% ...
$^{12}$ Universit{\'e} Libre de Bruxelles, Science Faculty CP230, B-1050 Brussels, Belgium \\
% ...
$^{17}$ Department of Physics, Loyola University Chicago, Chicago, IL 60660, USA \\
$^{18}$ Dept. of Physics and Astronomy, University of Canterbury, Private Bag 4800, Christchurch, New Zealand \\
% ...
$^{22}$ Niels Bohr Institute, University of Copenhagen, DK-2100 Copenhagen, Denmark \\
$^{23}$ Dept. of Physics, TU Dortmund University, D-44221 Dortmund, Germany \\
% ...
$^{26}$ Erlangen Centre for Astroparticle Physics, Friedrich-Alexander-Universit{\"a}t Erlangen-N{\"u}rnberg, D-91058 Erlangen, Germany \\
% ...
$^{40}$ Dept. of Physics and Wisconsin IceCube Particle Astrophysics Center, University of Wisconsin{\textendash}Madison, Madison, WI 53706, USA \\
% ...
$^{42}$ Department of Physics, Marquette University, Milwaukee, WI, 53201, USA \\
% ...
$^{44}$ Bartol Research Institute and Dept. of Physics and Astronomy, University of Delaware, Newark, DE 19716, USA \\
% ...
$^{63}$ Deutsches Elektronen-Synchrotron DESY, Platanenallee 6, 15738 Zeuthen, Germany  \\
$^{64}$ Institute of Physics, Sachivalaya Marg, Sainik School Post, Bhubaneswar 751005, India \\
% ...

\subsection*{Acknowledgements}

\noindent
The authors gratefully acknowledge the support from the following agencies and institutions:
USA {\textendash} U.S. National Science Foundation-Office of Polar Programs,
U.S. National Science Foundation-Physics Division,
U.S. National Science Foundation-EPSCoR,
Wisconsin Alumni Research Foundation,
Center for High Throughput Computing (CHTC) at the University of Wisconsin{\textendash}Madison,
Open Science Grid (OSG),
Advanced Cyberinfrastructure Coordination Ecosystem: Services {\&} Support (ACCESS),
Frontera computing project at the Texas Advanced Computing Center,
U.S. Department of Energy-National Energy Research Scientific Computing Center,
Particle astrophysics research computing center at the University of Maryland,
Institute for Cyber-Enabled Research at Michigan State University,
and Astroparticle physics computational facility at Marquette University;
Belgium {\textendash} Funds for Scientific Research (FRS-FNRS and FWO),
FWO Odysseus and Big Science programmes,
and Belgian Federal Science Policy Office (Belspo);
Germany {\textendash} Bundesministerium f{\"u}r Bildung und Forschung (BMBF),
Deutsche Forschungsgemeinschaft (DFG),
Helmholtz Alliance for Astroparticle Physics (HAP),
Initiative and Networking Fund of the Helmholtz Association,
Deutsches Elektronen Synchrotron (DESY),
and High Performance Computing cluster of the RWTH Aachen;
Sweden {\textendash} Swedish Research Council,
Swedish Polar Research Secretariat,
Swedish National Infrastructure for Computing (SNIC),
and Knut and Alice Wallenberg Foundation;
European Union {\textendash} EGI Advanced Computing for research;
Australia {\textendash} Australian Research Council;
Canada {\textendash} Natural Sciences and Engineering Research Council of Canada,
Calcul Qu{\'e}bec, Compute Ontario, Canada Foundation for Innovation, WestGrid, and Compute Canada;
Denmark {\textendash} Villum Fonden, Carlsberg Foundation, and European Commission;
New Zealand {\textendash} Marsden Fund;
Japan {\textendash} Japan Society for Promotion of Science (JSPS)
and Institute for Global Prominent Research (IGPR) of Chiba University;
Korea {\textendash} National Research Foundation of Korea (NRF);
Switzerland {\textendash} Swiss National Science Foundation (SNSF);
United Kingdom {\textendash} Department of Physics, University of Oxford.